\documentclass[12pt]{iopart}

\usepackage{graphicx}
\usepackage{cite}

\begin{document}

\title{Integrated Optical Dipole Trap for Cold Neutral Atoms with an Optical Waveguide Coupler}

\author{J Lee$^{1}$, D H Park$^{2}$, S Mittal$^{2}$, M Dagenais$^{2}$ and S L Rolston$^{1}$}

\address{$^{1}$Joint Quantum Institute, Department of Physics, University of
Maryland and National Institute of Standards and Technology, College Park, Maryland 20742, USA
$^{2}$Department of Electrical and Computer Engineering, University of Maryland, College Park, Maryland 20742, USA}

\ead{neoscien@umd.edu}

\begin{abstract}
 An integrated optical dipole trap uses two-color (red and blue-detuned)
traveling evanescent wave fields for trapping cold neutral atoms.  To achieve longitudinal confinement, we propose using  an integrated optical waveguide coupler, which provides a potential gradient along the beam propagation direction sufficient to confine atoms. This integrated optical dipole trap can support an atomic ensemble with a large optical depth due to its small mode area. Its quasi-$TE_{0}$ waveguide mode has an advantage over the $HE_{11}$ mode of a nanofiber, with little inhomogeneous Zeeman broadening at the trapping region. The longitudinal confinement eliminates the need for a 1-D optical lattice, reducing collisional blockaded atomic loading, potentially producing larger ensembles. The waveguide trap allows for scalability and integrability with nano-fabrication technology. We analyze the potential performance of such integrated atom traps.
\end{abstract}

\maketitle

%%%%%%%%%%%%%%%%%%%%%%%%%%  body  %%%%%%%%%%%%%%%%%%%%%%%%%%
\section{Introduction}

Optically trapped neutral atoms have been used for precision quantum
metrology \cite{Ye08} and quantum information processing \cite{Brennen99,Porto03}
because a neutral atom is an excellent frequency reference and atoms trapped in a far-off resonant optical field have a long
coherence time and a long lifetime. There is a wide variety of free-space configurations that have been employed to optically confine neutral atoms.  Free-space optical dipole traps \cite{Grimm00}
confine a cigar-shaped atomic cloud with a tightly focused traveling
wave, creating harmonic potentials along the propagation direction and
transverse directions. To enhance longitudinal confinement, a 1-D standing wave potential (optical lattice) \cite{Jessen96} can be used, creating an array of small-volume trapping sites. Atoms trapped in 2-D or 3-D optical lattices
are exploited to model quantum many-body systems in periodic potentials
and study fundamental condensed-matter problems \cite{Greiner02,Orzel01,Regal04}.
The two-color evanescent light surface trap by total internal reflection at the surface of a dielectric material was proposed in 1991 \cite{Ovchinnikov91} and realized in 2003 \cite{Hammes2003,Colombe03} on the surface of a prism.  Recent work trapping atoms with light guided in optical fibers includes a hollow-core photonic crystal fiber dipole trap \cite{Christensen08,Bajcsy09} and nanofiber 1-D optical lattices with two-color evanescent fields \cite{Sague07,Vetsch10}. An addition to trapping atoms, guided optical modes are also being investigated for trapping and manipulating nanoparticles and biomolecules in microfluidic slot-waveguide geometries \cite{Yang09} and in the evanescent field of tapered optical fibers \cite{Skelton12}.
%\cite{Sague07,Sague08,Stiebeiner10,Vetsch10,Vetsch12}.

Advances in optical waveguide technology suggest that developing atom traps based on waveguides may provide an interesting platform for quantum information processing and sensing, building on the scalability inherent in waveguide fabrication that may allow networks and integration that would be difficult with free-standing optical fibers. Using optical waveguides along with 1-D optical lattices has been proposed to trap
neutral atoms with two-color evanescent fields \cite{Burke02,Fu07}. Here we propose a design that provides longitudinal confinement without resorting to 1-D lattices, creating a trap similar to the workhorse optical dipole trap, but integrated on a surface rather than in free space.

The integrated optical dipole trap (IODT) with an integrated optical
waveguide coupler (IOWC) creates a small mode area resulting in a potentially large
optical depth of an atomic ensemble, which determines e.g., the performance
of quantum memory based on electromagnetically induced transparency
(EIT) \cite{Lukin03,Fleischhauer00,Fleischhauer05} or the phase sensitivity
of atomic spin polarization spectroscopy \cite{Smith03,Deutsch10}.
The small mode volume allows for  a low optical power to trap atoms. In
addition, the high optical depth creates a strong atom-light coupling,
and the IODT with an integrated  Bragg-grating cavity may additionally increase the optical depth
according to its cavity finesse. A nanofiber-based Bragg-grating
cavity has been explored and demonstrated \cite{Nayak11,Wuttke12},
 an interesting cavity QED (quantum electrodynamics)
system \cite{Kien09,Kien10} nearing the single-atom strong coupling regime.
Cavity QED with an atomic mirror created with  periodically trapped
atoms was also proposed \cite{Cheng12}. The IODT can operate with the same laser wavelengths as a nanofiber trap to allow
 a state-insensitive trap (for $^{133}$Cs)
that can improve the lifetime and coherence time of trapped atoms
\cite{Kien04a,Lacrote12,Goban12}. 

The IODT with two-color traveling evanescent
wave fields uses differential decay lengths of a blue and a red-detuned evanescent fields to trap atoms along the transverse direction similar to the nanofiber's radial atom confinement. In addition, the advantage of geometrical design of the IODT, such as an integrated optical waveguide coupler, creates a sufficient potential gradient along the propagation direction, which can confine atoms along that direction, obviating the need for a standing wave which is required for the axial confinement of the nanofiber atom trap. This should improve the loading efficiency over a nanofiber trap,  where  the optical lattice site volume results in inefficient, collisionally blockaded atomic loading \cite{Schlosser02} of 0 or 1 atom per lattice site. 

%%%%%%%%%%% newly revised %%%%%%%%%%%%%%%%%%
 In addition, using the  quasi-$TE_{0}$
mode of the IODT trap, we can make a stable trap perpendicular to the input polarization, where there is no longitudinal $E_{z}$ component and no ellipticity at the trapping region.
Nanofiber atom traps with the $HE_{11}$ mode usually create radial atom confinement using azimuthal symmetry breaking in the strongly guided regime where there exists more evanescent field parallel to the input polarization, but it has a non-negligible longitudinal component at its trapping region.
%%%%%%%%%%%%%%%%%%%%%%%%%%%%%%%%%%%%%%%%
 
The IODT leveraged by nano-fabrication technology
has the possibility of being scalable and integrated for a large
scale quantum memory and multiple arrays of atomic magnetometry sensors. The structural robustness and heat transfer of an IODT are much
better than the nanofiber because of its substrate and nano-fabricated
structure, allowing for higher optical powers and deeper traps.

\section{Trapping Atoms}

The IODT creates optical potentials that confine atoms along the transverse directions $(x,y)$ with the differential decay lengths of two-color traveling evanescent wave fields and confine atoms along the propagation direction $(z)$ with a sufficient potential gradient created by the  geometrical design of the IODT. The IODT optical trap works in a similar manner as an optical dipole trap. For a single atomic transition ($\omega_{0}$), the induced light shift (AC Stark shift) creates an optical
potential. This potential \cite{Grimm00} for a large detuning ($\Delta=\omega-\omega_{0}$) is
\begin{equation}
U_{opt}(\mathbf{r})=\frac{3\pi c^2}{2\omega_{0}^{3}}\frac{\Gamma}{\Delta}I(\mathbf{r}),
\end{equation}
where $I(\mathbf{r})$ is the laser intensity, and $\Gamma$ is the spontaneous decay rate of the excited state. A blue-detuned trapping beam creates a repulsive potential ($\Delta_{blue}>0$), and a red-detuned trapping beam produces an attractive potential ($\Delta_{red}<0$). For trapping near  ($\sim150$\,nm) the surface of a dielectric, the blue-detuned beam needs to compensate the attractive van der Waals potential (the Casimir-Polder interaction is negligible around the trapping region \cite{Hoffman11}). Approximating the surface to be an infinite dielectric, the van der Waals potential  is 
\begin{equation}
U_{vdW}(y)=-\frac{\epsilon-1}{\epsilon+1}\frac{C^{(3)}}{y^{3}}=-\frac{C_{vdW}}{y^{3}},
\end{equation}
where $\epsilon$ is the dielectric permittivity, $y$ is the distance from the waveguide surface, and $C^{(3)}$ is determined from atomic dipole transition \cite{Burke02,Aspect00}. %\cite{Hoffman11,Courtois96,Landragin96}.
The total  potential along the vertical direction of the waveguide is
\begin{eqnarray}
U_{tot}&=&U_{blue}+U_{red}+U_{vdW} \\
&=&\frac{3\pi c^2}{2\omega_{0}^3}\frac{\Gamma}{\Delta_{blue}}I_{blue}(\mathbf{r})+\frac{3\pi c^2}{2\omega_{0}^3}\frac{\Gamma}{\Delta_{red}}I_{red}(\mathbf{r})-\frac{C_{vdW}}{y^3}.\nonumber
\end{eqnarray}
If the chosen wavelengths are near to other atomic transitions, the optical potential can be generalized by summing the light shifts for each laser over each transition. We use as an example optical trapping of cold $^{87}$Rb atoms in their ground-state manifold 5S$_{1/2}$ with two excited-state manifolds of 5P$_{1/2}$ and 5P$_{3/2}$ as 
\begin{equation}
\frac{\Gamma}{\Delta} = \frac{1}{3}\frac{\Gamma_{D_{1}}}{\Delta_{D_{1}}} + \frac{2}{3}\frac{\Gamma_{D_{2}}}{\Delta_{D_{2}}}.
\end{equation}
We can trap cold neutral atoms using  single blue-detuned and single red-detuned traveling evanescent waves of the IODT, as the IODT has a potential gradient along the propagation direction similar to an optical dipole trap without standing waves. The IODT design can easily be generalized for trapping other atomic species, with dimensions appropriately scaled to the relevant trapping laser wavelengths.

\section{Vector Light Shift of Trapping Beams}
% This section is newly added for revision.%

The atom-light interaction is generally described with scalar, vector, and tensor light shifts \cite{Deutsch10}. 
\begin{eqnarray}
\hat{H}_{ls} &=& \hat{H}_{0} + \hat{H}_{1} + \hat{H}_{2} \\
&=& \sum_{n' J' F'} V_{0, J J'} \{ C^{(0)}_{J'F'F} |\vec{\epsilon}|^2 + i C^{(1)}_{J'F'F}(\vec{\epsilon}^{*} \times \vec{\epsilon}) \cdot \mathbf{F} + C^{(2)}_{J'F'F} (|\vec{\epsilon} \cdot \mathbf{F}|^2 - \frac{1}{3}\mathbf{F}^2|\vec{\epsilon}|^2) \} \nonumber,
\end{eqnarray}
where $V_{0, J J'} = -\frac{1}{4} \alpha_{0, J J'} {|E_0|}^2 = \frac{3\pi c^2}{2\omega_{J' J}^3}\frac{\Gamma_{J' J}}{\Delta_{F' F}}I(\mathbf{r})$,  $\vec{\epsilon}$ is a light polarization, $\mathbf{F}$ is the total angular momentum operator, the tensor coefficients $C^{(K)}$ are calculated from Wigner-Eckart theorem, an $\alpha_{0, J J'}$ is the characteristic polarizability of atomic transition from $n J$ (lower level) to $n' J'$ (upper levels). The scalar light shift ($\hat{H}_{0}\propto \frac{I(\mathbf{r})}{\Delta_{F' F}}$) yields the  trapping of atoms with an optical dipole trap.  Vector ($\hat{H}_{1}$) and tensor ($\hat{H}_{2}$) light shifts depend on the polarization state of light and populated magnetic sublevels  for a given quantization axis. For example, when the propagation direction of light  and the populated atomic state $|F,m_{F}\rangle$ are colinear along the quantization ($\hat{z}$) axis, the vector light shift ($\hat{H}_{1} \propto C^{(1)} S_{3} F_{z}$, where $S_{k}$ is Stokes vector components \cite{Jackson}) is proportional to the ellipticity of the light. For  nanofiber atom traps, the suppression of vector light shifts from trapping beams is important due to the  longitudinal component at the trapping region. The vector light shift  induces a state-dependent light shift and a fictitious magnetic field, which can make magnetometry and Faraday rotation measurements challenging.  State-insensitive traps \cite{Kien04a,Lacrote12} can have long lifetimes and coherence times, by matching the potential curvatures of the ground and excited state atoms (including Zeeman sublevels)  with a properly chosen two-color magic wavelength trap. 

For the ground state ($n S_{1/2}$) of alkali atoms in the far-detuned regime where $\Delta_{FF'} \gg \delta_{F'}$ (excited-state hyperfine splitting), the sums of tensor light shifts are cancelled for $D_{1}$ and $D_{2}$ transitions. The sums of tensor light shifts from the excited-state hyperfine transition to all upper transitions do not completely cancel, but are an order of magnitude smaller than vector light shifts, and thus  can be neglected. 

A state-insensitive, compensated nanofiber trap was proposed and demonstrated for $^{133}Cs$ atoms with a two-color magic wavelength atom trap and the cancellation of vector light shifts of a forward propagating blue-detuned light by means of adding a backward propagating off-resonant blue-detuned beam \cite{Lacrote12,Goban12}. In the nanofiber's strongly guided regime ($a_{radius}<\lambda/2$), a quasi-linear $HE_{11}$ mode's electric field parallel to the input polarization generates larger evanescent fields, which a nanofiber-based 1D lattice  exploits, but has a non-negligible longitudinal component ($E_{z}$) due to the discontinuity of electric field and azimuthal symmetry breaking.   The orthogonal linear polarization of blue and red fields has been also demonstrated in a nanofiber \cite{Vetsch10}. The trapped atoms in the hyperfine ground states experience no longitudinal component along the radial and axial directions, but the trapping potential in the  azimuthal direction experiences vector light shifts and  inhomogeneous Zeeman broadening \cite{Goban12}. 

In the case of the rectangular waveguide (see Fig. \ref{fig_IOWC_Cross-Section}), the fundamental mode is  a quasi-$TE_{0}$ mode, which also has an electric field along the input polarization like the $HE_{11}$ mode. However, unlike $HE_{11}$ mode, the waveguide has a large evanescent field perpendicular to the input polarization because of its thin height in addition to a higher index of refraction. Therefore, we can have optical potentials of blue and red-detuned light along the direction perpendicular to the input polarizations with moderate optical powers. Furthermore, the waveguide can decouple the azimuthal noise in the trap potential compared to the nanofiber traps \cite{Vetsch10,Goban12}, and inhomogeneous Zeeman broadening and the vector light shift is  suppressed.

 Waveguide atom trapping with the quasi-$TE_{0}$ mode has the advantage of virtually no vector light shift at the atom trapping region because the quasi-$TE_{0}$ mode's evanescent field perpendicular to the input polarization has a very small  $E_{z}$ component. At the potential minimum, the waveguide mode has   $|E_{z}|^2/|E|^2$ of less than 0.3\,\%, ten times less than  a nanofiber trap.   Even in the motional ground state of the transverse direction (x-axis, assuming a harmonic trap with the trap frequency $\omega_{x}=$2$\pi\cdot$80\,kHz) the atoms can experience an inhomogeneous broadening due to the vector light shift, but for reasonable trapping parameters, the broadening is small (0.1\,\%). The longitudinal component mostly comes from two side edges of the waveguide cross-section, and may be further decreased  by redesigning the waveguide's width within the condition of single-mode operation.

\section{Waveguide Design}

The optical waveguide is designed for efficient atom trapping with
low insertion loss at both operating wavelengths (760\,nm and 1064\,nm for $^{87}$Rb atoms),
while satisfying straightforward operation and noncritical design
criteria that facilitate the design/fabrication process. Various types
of integrated optical waveguide structures, such as rib, ridge, pedestal,
strip, trapezoid, and nanoslot, can be used for the atom trapping.
Here, CMOS-compatible Si$_{3}$N$_{4}$ strip-loaded optical waveguide
on silicon on insulator (SOI) is considered because of relatively
low loss at the operating wavelengths, low coupling loss, and a large evanescent field caused by weak optical guiding, compared
to Si or GaAs waveguides. 

\begin{figure}
\centering\includegraphics[width=1\textwidth]{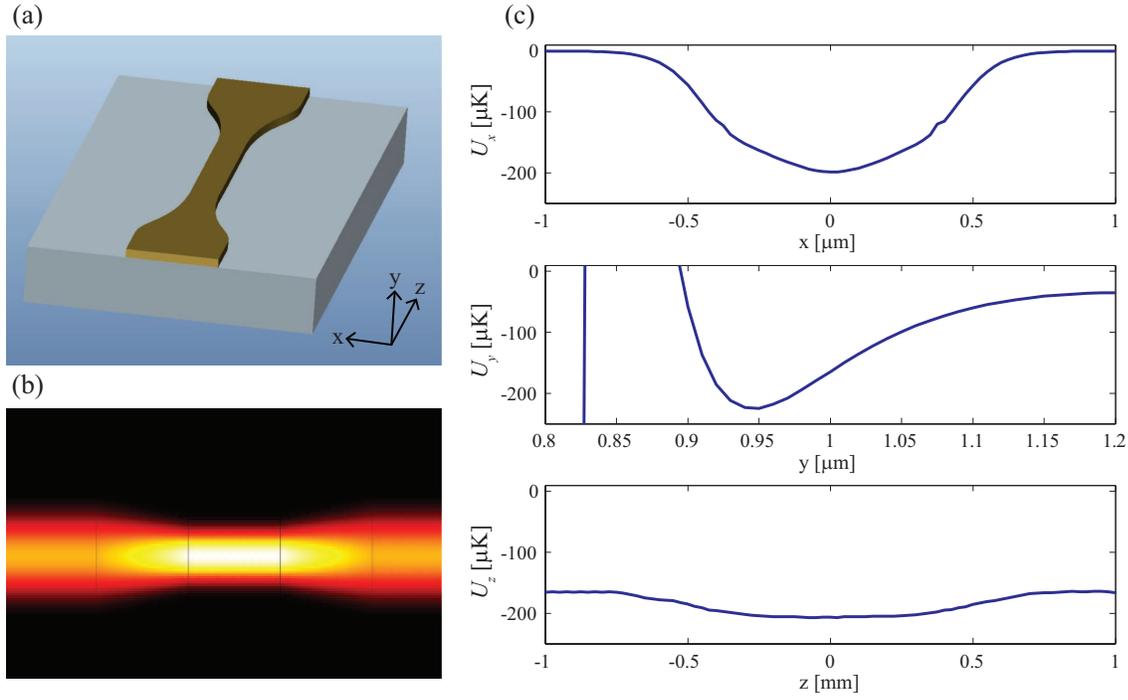}
\caption{(a) Hourglass strip-loaded optical waveguide (3-D plot, not to scale) (b) 760\,nm field distribution (xz plane, not to scale) (c) Trapping potentials for $^{87}$Rb atoms ($P_{760}\simeq12\,\mathrm{mW}$ and $P_{1064}\simeq40\,\mathrm{mW}$); $U_{x}$ (y $\simeq$ 0.95\,$\mu$m) and $U_{y}$ (x $\simeq$ 0\,$\mu$m) with $U_{vdW}$ are potentials along the transverse direction and $U_{z}$ (x $\simeq$ 0\,$\mu$m, y $\simeq$ 0.95\,$\mu$m) are a potential along the propagation
direction. The evanescent field along y-axis starts at 0.8\,$\mu$m. }
\label{fig_STOW}
\end{figure}

%\caption{(a) Hourglass strip-loaded optical waveguide (3-D plot, not scaled) (b) 760\,nm field distribution (xz plane) (c) The trapping potentials for $^{87}$Rb atoms with  $P_{760}\simeq12\,\mathrm{mW}$ and $P_{1064}\simeq40\,\mathrm{mW}$ are $(U_{x},U_{y},U_{z})\:\simeq(-203.5,-226,-44)\,\mathrm{\mu K}$, where $U_{x}$ and $U_{y}$ with $U_{vdW}$ are potentials along the transverse direction and $U_{z}$ are a potential along the propagation direction.}

One may think of an hourglass shape of strip-loaded optical waveguides (see Fig. \ref{fig_STOW} (a)) in order to reduce the optical potential at the ends to provide longitudinal confinement. This does not, however, create a sufficient potential gradient  along the propagation direction to achieve the desired device performance (see Fig. \ref{fig_STOW} (c)). Simply splitting each input and output port into two optical waveguides,  a well-known $2\times2$ waveguide coupler configuration, allows a much larger extinction ratio along the propagation direction and a 3\,dB power reduction in each waveguide caused by spatial separation of optical fields into two transverse directions. Smaller separation
between the ports is better for adiabaticity, but separation
of less than 1\,$\mu$m leads to an insufficient potential gradient along
the propagation direction due to the residual light between two ports.
High refractive index material such as silicon or III-V compounds might be used
in the form of a strip-loaded waveguide or a nanoslot \cite{Almeida04},
which requires a large index difference to induce intense evanescent
fields. However, special care should be taken to choose an operating
wavelength and its polarization for both low optical loss and no ellipticity. Organic
polymer materials are also good candidates because of their intrinsic
properties such as low optical loss, weak dispersion, and low refractive index. Polymers
can be molecularly engineered to have nonlinearity for active device
applications controlled by electrical and optical signals, which may allow for additional device functionality.

\begin{figure}
\centering\includegraphics[width=1\textwidth]{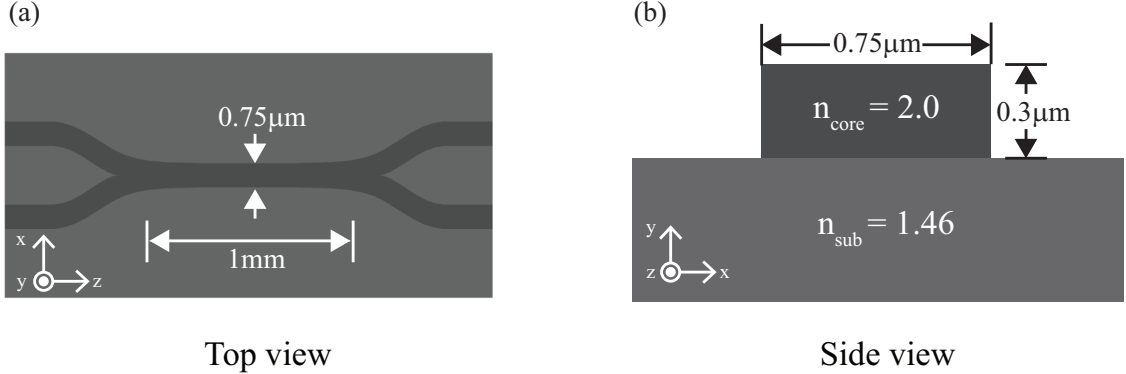}
\caption{An integrated optical waveguide coupler (a) Top view of an optical waveguide coupler (xz plane, not to scale). (b) Cross-sectional view of an optical waveguide coupler (xy plane, not to scale).}
\label{fig_IOWC}
\end{figure}

The  waveguide is constructed with a silicon nitride core (Si$_{3}$N$_{4}$,
n$_{\mathrm{core}}=2.05$) and a silicon oxide substrate (SiO$_{2}$, n$_{\mathrm{sub}}\simeq$1.46) as shown in Fig. \ref{fig_STOW}. Si$_{3}$N$_{4}$ can be grown by various types
of deposition techniques, such as plasma enhanced chemical vapor deposition (PECVD) and  inductively coupled plasma (ICP), and its refractive index
is controllable in the range of $1.8\sim2.5$. The core size of the waveguide
at the center of the IOWC is 750\,nm\,$\times$\,300\,nm (see Fig. \ref{fig_IOWC}). The length of the core waveguide is designed to be 1\,mm, but may be longer depending on applications and lithographical capability.

The waveguide is designed for atom trapping and single mode operation so that only the quasi-$TE_{0}$ mode can
propagate along the waveguide at both operating wavelengths. For this
design, we consider how to obtain an intense evanescent field while maintaining
the single mode operation and high optical coupling efficiency. As
the height of waveguide decreases, more evanescent field can be obtained
for a small mode area, resulting in lower power requirements for the same depth trap.  However, when the height of the IOWC
is smaller than $\sim250$\,nm, the ideal convergence condition of the 1064\,nm
waveguide modes is not satisfied. We choose  a 300\,nm thick waveguide to allow $\pm\,50$\,nm
design uncertainty. The width of the waveguide can be similarly treated.
As the width diminishes, the evanescent field intensifies due to the
reduced mode area. We note that improper height and/or width
can cause multi-mode interference (MMI) or poor optical coupling efficiency.
Our waveguide design is optimized for guiding 760\,nm and 1064\,nm
trapping lasers (see Fig. \ref{fig_IOWC}). We choose
a 750\,nm width to prevent MMI in the IOWC. This sub-micron width design needs only a few mW
of optical power of 760\,nm  and 1064\,nm trapping light to produce useable trap depths because of the small
cross-sectional area. In contrast, a few $\mu$m beam diameter of an
optical dipole trap requires a few hundreds of mW of optical power for
the same blue- and red-detuned trapping lasers due to its large 
mode area. Of course, the width of the IOWC can be adjusted to vary
the trapping volume. For a large trapping volume, which may be useful to increase the number of trapped atoms, we may need to increase
optical powers and to consider the MMI condition carefully. The IOWC is formed to
include an adiabatic region in the coupling sections for a single mode
operation and low bending loss.

\section{Simulations}

\begin{figure}
\centering\includegraphics[width=1\textwidth]{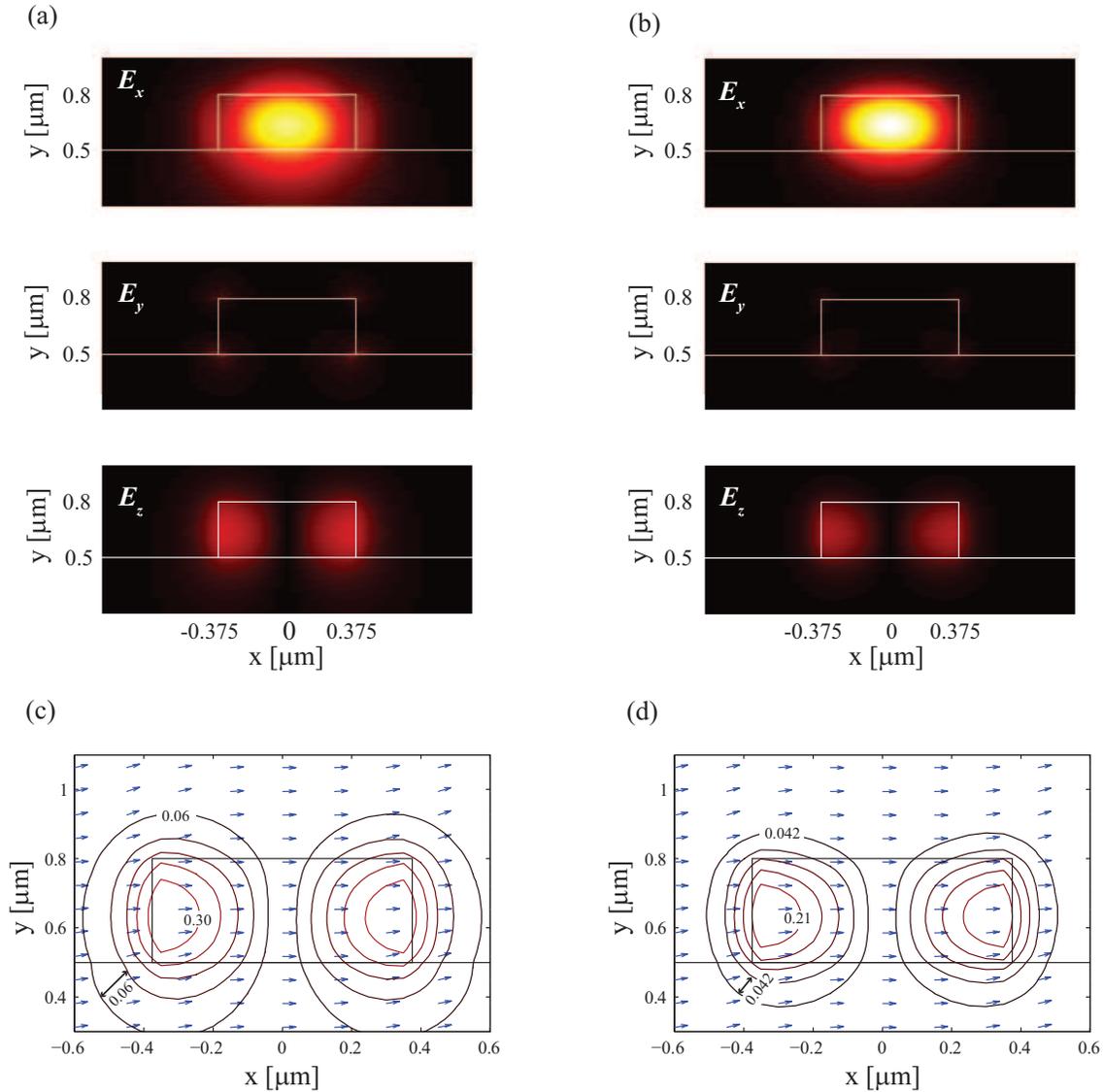}
\caption{Cross-sectional waveguide modes and quasi-$TE_{0}$ mode's polarization (a) $E_{x}$, $E_{y}$, $E_{z}$ waveguide modes of 1064\,nm light (xy plane); the input polarization along the x-axis and the evanescent red-detuned trapping field perpendicular to the input polarization.  (b) $E_{x}$, $E_{y}$, $E_{z}$
waveguide modes of 760\,nm light (xy plane); the input polarization along the x-axis and the evanescent blue-detuned trapping field perpendicular to the input polarization.  (c) 1064\,nm waveguide mode's polarization with $E_{z}$ mode's contour plot (xy plane); 0.30 means 30\% of 1064\,nm mode's max. (d) 760\,nm waveguide mode's polarization with $E_{z}$ mode's contour plot (xy plane); 0.21 means 21\% of 760\,nm mode's max.}
\label{fig_IOWC_Cross-Section}
\end{figure}

\begin{figure}
\centering\includegraphics[width=1\textwidth]{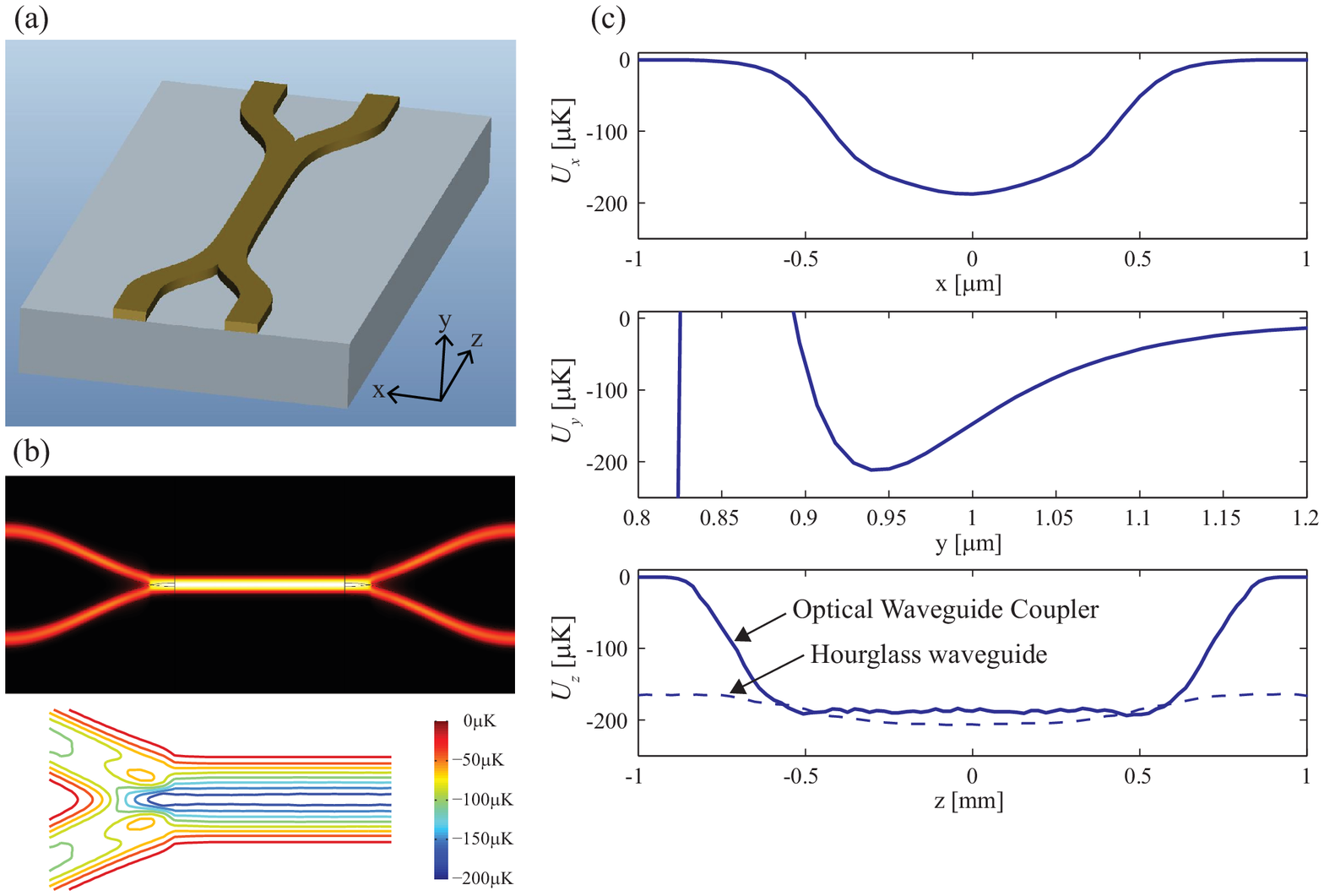}
\caption{An integrated optical waveguide coupler (a) Optical waveguide coupler
(3-D plot, not to scale) (b) top: 760\,nm field distribution (xz plane, not to scale), bottom: contour plot of trapping potential (y $\simeq$ 0.95\,$\mu$m) around the junction (xz plane, not to scale) (c) Trapping potentials for $^{87}$Rb atoms ($P_{760}\simeq12\,\mathrm{mW}$ and $P_{1064}\simeq40\,\mathrm{mW}$); $U_{x}$ (y $\simeq$ 0.95\,$\mu$m) and $U_{y}$ (x $\simeq$ 0\,$\mu$m) with $U_{vdW}$ are potentials along the transverse direction and $U_{z}$ (x $\simeq$ 0\,$\mu$m, y $\simeq$ 0.95\,$\mu$m) are a potential along the propagation direction. The evanescent field along y-axis starts at 0.8\,$\mu$m. The apparent roughness in the potentials is due to numerical error. The dotted line is the $U_{z}$ of an  hourglass strip-loaded optical waveguide (see Fig. \ref{fig_STOW} (c)).}
\label{fig_IOWC_Utrap}
\end{figure}

We simulated the 3-D structure of the IODT with the IOWC using FIMMWAVE based on the FMM
(Film Mode Matching) method that numerically calculates a 2-D mode along
the beam-propagation direction \cite{Sudbo93a,Sudbo93b}. The 2D cross-sectional
views of two waveguide modes (760\,nm and 1064\,nm) that can trap
atoms near the surface  are shown in  Fig. \ref{fig_IOWC_Cross-Section}. The blue-detuned potential (760\,nm) is required to compensate the attractive van der Waals potential. The different field decay lengths of 760\,nm and 1064\,nm evanescent
fields create an optical potential to trap atoms at a distance of
$\sim150$\,nm above the waveguide (see $U_{y}$ of Fig. \ref{fig_IOWC_Utrap}(c)).

The waveguide designed for single mode operation has only the quasi-$TE_{0}$ mode at both operating wavelengths. The quasi-$TE_{0}$ mode contains strong
$E_{x}$ (horizontal electric field) and weak $E_{y}$ and $E_{z}$ modes
around the corners of the waveguide as shown in Fig. \ref{fig_IOWC_Cross-Section}.
Therefore, at the region where the atoms are trapped, $\sim150$\,nm above
the central surface of IOWC, there exists negligible $E_{y}$ and $E_{z}$ fields.
Compared to the nanofiber trap using the $HE_{11}$ mode's electric field parallel to the input polarization, the proposed device
has a negligible  polarization ellipticity (see Fig. \ref{fig_IOWC_Cross-Section} (c), (d)) at the trapping
area along the direction perpendicular to the input polarization.

%\caption{An integrated optical waveguide coupler (a) Optical waveguide coupler (3-D plot, not scaled) (b) 760\,nm field distribution (xz plane) (c) The trapping potentials for $^{87}$Rb atoms with $P_{760}\simeq12\,\mathrm{mW}$ and $P_{1064}\simeq40\,\mathrm{mW}$ are $(U_{x},U_{y},U_{z})\:\simeq(-192,-214,-190)\,\mathrm{\mu K}$, where $U_{x}$ and $U_{y}$ with $U_{vdW}$ are potentials along the transverse direction and $U_{z}$ are a potential along the propagation direction. The apparent roughness in the potentials is due to numerical error. The dotted line is the $U_{z}$ of an hourglass strip-loaded optical waveguide (See Fig. \ref{fig_STOW} (c)).}

The trapping potentials along the transverse ($x$ and $y$) and the
propagation ($z$) directions are calculated (see Fig. \ref{fig_IOWC_Utrap}(c)).
The  optical potential gradient confines atoms along
the propagation direction. Nanofiber-based traps require  standing waves for confining atoms along the propagation direction,
but the IODT with the IOWC can use only two-color traveling evanescent
waves due to its geometry. The trapping potentials ($U_{x}$, $U_{y}$,
$U_{z}$) at the central trapping region of the waveguide are about $-200\,\mathrm{\mu K}$
(see Fig. \ref{fig_IOWC_Utrap}) for $P_{760}\simeq12\,\mathrm{mW}$ and $P_{1064}\simeq40\,\mathrm{mW}$.
With a minimal height of the IOWC ($\sim200$\,nm), the required optical
powers of trapping beams can be further reduced by a factor of two, but we chose the values considering the convergence of the waveguide
modes. Trapping frequencies can be estimated using the harmonic trap
approximation as $\omega_{i\,(=x,y)}= \sqrt{\frac{4U_{0}}{m w_{0,i}^{2}}}$. The trap frequencies along the transverse direction are $(\omega_{x},\omega_{y})\simeq2\pi\cdot(80\,\mathrm{kHz},493\,\mathrm{kHz})$ (see Fig. \ref{fig_IOWC_Utrap}).

\begin{figure}
\centering\includegraphics[width=0.6\textwidth]{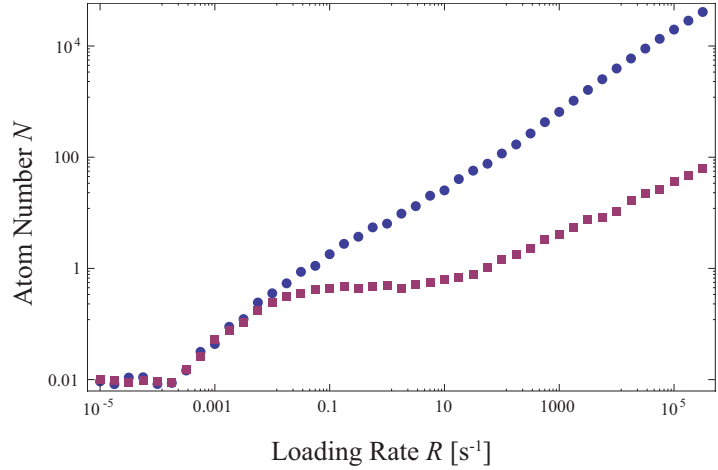}
\caption{Collisionally blockaded atomic loading (Monte-Carlo simulations with
$20,000$ samples for each loading rate); the IODT with the IOWC has
$\beta^{'}=0.01\,\mathrm{s^{-1}}$ and $\Gamma=0.02\,\mathrm{s^{-1}}$ (blue dots), and the nanofiber-based
1-D optical lattices have $\beta^{'}=28.2\,\mathrm{s^{-1}}$ and $\Gamma=0.02\,\mathrm{s^{-1}}$
(red squares).}
\label{fig_Collisional_Blockade}
\end{figure}

The IODT with no required 1-D optical lattice can eliminate the collisional blockade
regime that limits the average number of trapped atoms per lattice site to 0.5 (see Fig. \ref{fig_Collisional_Blockade}). (Collisional blockade arises from two-body collisions that create very high atom-loss rates for lattice sites with more than one atom. 
%newly added for revision% 
A recent experiment \cite{Grunzweig10} shows that  filling factor can be improved to 0.8 beyond the 0.5 limit using blue-detuned light-assisted collisions.) 
%%%%%%%%%%%%%%%%%%%%
We expect that an IODT could sustain more atoms than the nanofiber-based 1-D optical lattices due
to reduced two-body collisions, but the number the trapped atoms needs
to be carefully determined in the experiment. The decay process of
the trapped atoms is modeled with atomic loading rate ($R$), the
background gas collision ($\Gamma$), and light-assisted two body
collisions ($\beta'=\beta/V_{eff}$) as follows:
\begin{equation}
\frac{dN}{dt}=R-\Gamma N-\beta'N(N-1).
\end{equation}
We estimated the light-assisted two-body collision rate for the IODT
with the IOWC ($\beta'\sim0.01\,\mathrm{s^{-1}}$) and the nanofiber-based 1-D optical
lattices ($\beta'\sim28.2\,\mathrm{s^{-1}}$) considering an effective trap volume,
$V_{eff}$ \cite{Wallace92}. The atomic loading process from magneto-optically
trapped (MOT) atoms to an optical dipole trap needs a near-resonant MOT
beam to provide cooling in the potential. This produces excited-state atoms, and thus light-assisted two body collisions.
Dominated by a radiative escape process, the collision rate depends on the intensity, detuning,
and duration of the near-resonant beam \cite{Grunzweig10,Forster06,Fuhrmanek11}.
The trap volume of the IODT with the
IOWC is a few thousand times more than that of the 1-D optical lattice
site. Therefore light-assisted two body collisional loss is reduced
for the IODT, and we may avoid a collisionally blockaded loading process
\cite{Schlosser02,Sortais11}.  

The loading efficiency of the nanofiber-based optical lattices is
lower than that of the IODT case due to $\beta'$, but during an adiabatic
loading procedure, a recapturing process \cite{Forster06} by the
near-resonant MOT beam can enhance the atom loading into the 1-D array
of single atom traps. After turning off the adiabatic loading,
the single atoms trapped in the nanofiber optical lattices have no
more two-body or three-body collisions.  Without the presence of near-resonant light, the collision rate in an IODT will be dominated by three-body loss, but this should be negligible at expected densities.  A near-resonant
probe for an absorption measurement can induce light-assisted
collisions (two-body loss) and light-induced heating
of the atoms (effectively a one-body loss mechanism). Those could however be greatly suppressed
with an off-resonant probe ($\delta_{prob}>300\,\mathrm{MHz}$) for a phase-shift
measurement. A detailed comparison of atom number between  a nanofiber trap or an IODT with
the IOWC is somewhat difficult. The total number of trapped atoms
is determined by the initial density of MOT cloud, the efficiency
of adiabatic loading process from MOT atoms to the optical trap,
two-body collision rates with a given trapping volume, and the intensity,
detuning, and duration of the near-resonant MOT beam (or probe beam
for each measurement protocol).

%newly added for revision%

One additional possible advantage of the IODT over the nanonfiber trap is potentially reduced phase fluctuations. It is thought that phase fluctuations of the 1D lattice due to spontaneous Brillioun scattering limits the lifetime of trapped atoms on the nanofiber traps.  By eliminating the lattice, one eliminates the sensitivity to phase fluctuations. %%%%%%%%%%%%%%%%%%%%

\section{Conclusion}
An atomic ensemble with a large optical depth stored in an IODT provides a versatile and scalable tool for a variety of quantum information applications, including quantum memory, quantum repeaters \cite{Duan01} and magnetic sensors based on atomic spin polarization spectroscopy. As an example, EIT-based quantum memory in a $\Lambda$ system enables us to store a signal photon pulse in the collective atomic ensemble as a photonic dark state polariton \cite{Fleischhauer00}, using a classical control field that induces slow light, and later retrieve the signal photon from the atomic medium with the control field. The efficiency of the  quantum memory depends on the optical depth of the atomic medium. The IODT has a high optical depth due to the small-mode volume, and the output photon is automatically guided and hence can be used with high efficiency.  Therefore, we can use this as a scalable and integrable quantum memory unit.

%An atomic ensemble with a large optical depth stored in an IODT provides a versatile and scalable tool for a variety of quantum information applications, including quantum memory, quantum repeaters \cite{Duan01} and magnetic sensors based on atomic spin polarization spectroscopy. As an example, EIT-based quantum memory in a $\lambda$ system enables us to store a signal photon pulse into the collective atomic ensemble as a photonic dark state polariton \cite{Fleischhauer00} using a classical control field. A signal photon pulse resonant with the EIT window enters the atomic medium, and the classical control field induces slow light that results in spatial compression. When the signal photon is inside of the atomic medium, the control field is adiabatically reduced, bring the group velocity down to zero. After closing the EIT window, the signal photon is stored in the atomic medium. When the pulse needs to be retrieved, the control field is turned back on. The pulse then resumes its propagation and leaves the EIT medium. The efficiency of quantum memory depends on the optical depth of atomic medium. The IODT has a high optical depth due to the small-mode volume, and the output photon is automatically guided and hence can be used with high efficiency.  Therefore, we can use this as a scalable and integrable quantum memory unit.

As a second example, the waveguide can be used for atomic spin polarization spectroscopy
\cite{Smith03,Deutsch10}. The quasi-$TE_{0}$ mode's electric field perpendicular to the input polarization has no longitudinal $E_{z}$ component at the trapping region. Therefore, the trapping beams do not induce vector light shifts, and the vector light shift of a linearly polarized off-resonant probe for Faraday measurement is not limited by the trapping beam's vector light shift.
%%%%%%%%%%%%%% newly revised %%%%%%%%%%%%%%%%%%%%%%%%%

 %%%%%%%%%%%%%%%%%%%%%%%%%%%%%%%%%%%%%%%%%%%%%%%%%
Therefore, the IODT has the potential to be a compact and robust atomic magnetometry sensor, with less need to develop schemes to 
cancel vector light shifts.

The integrated optical dipole trap (IODT) with an integrated optical
waveguide coupler (IOWC) can trap cold neutral atoms near the surface
with two-color traveling evanescent wave fields. The proposed structure
of the  IOWC creates a confining potential along
the propagation direction without optical lattices or standing waves.  Because of the
small mode volume, the IODT can produce a high optical depth of an atomic
ensemble, the figure-of-merit for  the performance of quantum memory or an atomic
magnetometry sensor. The scalability and integrability of the IODT
are promising for a large scale quantum memory or a multiple-array
of atomic magnetometry sensors. The IODT eliminates collisionally
blockaded atomic loading and thus may have a better loading efficiency than
the nanofiber-based 1-D optical lattices. Furthermore, a quasi-$TE_{0}$
mode has no polarization ellipticity at the trapping region perpendicular to the input polarization, and vector light shift cancellation
\cite{Lacrote12,Goban12} is unnecessary. The IODT with the IOWC
can also be designed for magic wavelengths (such as $^{133}$Cs (935.7\,nm and 684.8\,nm)), where the trapped atoms can have a longer
coherence time and a longer lifetime. In addition, the structural
robustness and power tolerance of the IODT are ideal for deployment in an ultra-high vacuum environment.

\clearpage

%\section*{Acknowledgements}
%This research is funded by NSF-PFC and ARO Atomtronics MURI.

\end{document}